\documentstyle[aps,prl,twocolumn,psfig]{revtex}
\begin{document}


\draft
\preprint{\today}
\title{Fluctuation dissipation ratio in an aging Lennard-Jones glass}

\author{Jean-Louis Barrat }
\address{ D\'epartement de Physique des Mat\'eriaux \\
Universit\'e Claude Bernard and CNRS, 69622 Villeurbanne Cedex, France}

\author{Walter Kob }
\address{ Institut f\"ur Physik, Johannes Gutenberg-Universit\"at,
Staudinger Weg 7, D-55099 Mainz, Germany}

\date{\today}

\maketitle

\begin{abstract}
By using extensive Molecular Dynamics simulations, we have determined
the violation of the fluctuation-dissipation theorem in a Lennard-Jones
liquid quenched to low temperatures. For this we have calculated
$X(C)$, the ratio between a one particle time-correlation function $C$
and the associated response function. Our results are best fitted by
assuming that $X(C)$ is a discontinuous, piecewise constant function.
This is similar to what is found in spin systems with one step replica
symmetry  breaking. This strengthen the conjecture of a similarity
between the phase space structure of structural glasses and such spin
systems.

\end{abstract}



 
\narrowtext

Obtaining information on the phase space structure of glassy systems is
a very difficult challenge. By definition, relaxation times in a glass
are so long as to preclude equilibration within an experimental (or
numerical) time scale, except perhaps for very small systems
\cite{ParisiColuzzi,Heuer}. Exploration of phase space in these systems
is necessarily incomplete, and the results from any experimental
investigation cannot be expected to be representative of a well defined
statistical ensemble. Hence, although many conjectures have been
formulated concerning the structure of phase space in glassy systems
\cite{energy_landscape}, very little is actually known.

A promising route, that might to some extent bypass this intrinsic
difficulty, is the idea  that relevant information on phase space
structure is encoded in the nonequilibrium dynamics of glassy systems.
This idea was actively developed in the field of spin glasses
\cite{Vincent96,Bouchaud97,rieger}, but its extension to the field of
structural glasses is more recent
\cite{kob_barrat,nagel,parisi_prl,parisi_jpa}. Among the important
quantities that can be investigated in a nonequilibrium system is the
so called fluctuation-dissipation ratio $X$, defined in the following
way. Consider an observable $A$ whose normalized autocorrelation
function  will be denoted by $C$, and let $R$ be the response function
of $A$ to a field $H$ conjugate to $A$.  Then, for a system in
equilibrium at temperature $T$, $C(t)$ is related to the response
function $R(t)$ of the system to $H$ by the usual fluctuation
dissipation theorem (FDT), $R(t)= -{1\over k_B T} {dC\over dt}$.  In a
system that is out of equilibrium (e.g., a system that has been
quenched at $t=0$ to a low temperature) the property of time
translation invariance is lost, and $C$ and $R$ become functions of two
time variables, e.g. $C(t',t)= \langle A(t') A(t)\rangle$. A formal way
of generalizing the usual FDT consists in writing, for $t'>t$
\begin{equation}
R(t',t) = {1\over k_B T} X(t',t) {\partial C(t',t) \over \partial t} \quad ,
\label{e1} 
\end{equation} 
which in this form is merely a definition of $X$. The importance of
this ``FDT violation factor" $X(t',t)$ was recognized in the context of
mean field theories of spin glasses \cite{lcjk}, where it appears that
$X(t',t)$ has the properties discussed below. For this discussion it is
useful to consider the situation in which the system is driven out of
equilibrium at time $t=0$, then aged for a waiting time $t_w$ after
which the measurement of the time correlation functions
$C(t_w+\tau,t_w)$ are started. For mean field models exhibiting glassy
behavior it has been shown that in the limit of long times, $t_w,\tau
\rightarrow \infty$, $X(t_w+\tau,t_w)$ is a function of the correlation
function $C$ only, i.e.
\begin{equation}
 X(t_w+\tau,t_w)=x(C(t_w+\tau,t_w)),
\label{xC} 
\end{equation}
where $x$ is now a function of one variable.  In this asymptotic limit
two regimes can be distinguished. If $t_w$ is kept fixed,
$C(t_w+\tau,t_w)$ eventually becomes a function of $\tau$ only.  The
limiting value of this function for $\tau \rightarrow \infty$ is the
Edwards-Anderson parameter associated with observable $A$, which is
usually denoted by $q_{EA}$. Obviously $q_{EA}$ vanishes for an
equilibrium system, and differs from zero in a nonergodic system. For $
1>C>q_{EA}$, we have $x(C)=1$, meaning that the FDT holds. This means
that for time differences that are small compared to the waiting time
$t_w$, the response of the system is similar to that of an equilibrium
system, in spite of the fact that only a restricted part of phase space
is explored.  Nonequilibrium, or ``aging" features, show up in a
different limit, namely for $\tau > t_w$. In this limit, the
correlation function depends on both $t_w$ and $\tau$ in a  nontrivial
way, typically $C(t_w+\tau,t_w) = F(h(t_w+\tau)/h(\tau))$, where $h(x)$
is a monotonically increasing function. In this ``aging" regime,
$C<q_{EA}$, and $x(C)<1$. The system starts to sample a larger portion
of phase space, but this sampling is an out of equilibrium process, and
does not obey the equilibrium FDT.

An important property of $x(C)$, again discovered in the context of
mean-field spin glasses, is that the general structure of this
function  is identical to that of the function $x_{stat}(q)$ obtained
by inverting the integral of the Parisi function $q_{stat}(x)$. The
latter reflects the probability distribution of overlaps between
replicas of the same system, and does not involve any dynamical
consideration \cite{fisherhertz}. At present, the similarity between
these two functions has escaped physical understanding, although a
formal justification has recently been proposed
\cite{franzparisimezard}.  This similarity between the two functions
is, nevertheless, believed to be a general feature. If this is the
case, the implication is that the study of a dynamical quantity such as
$x(C)$  provides indirect information on the structure of phase space.
So far $x(C)$ was determined for the 3d Edwards-Anderson model
\cite{franzrieger}, for ferromagnetic coarsening \cite{abarrat}, for
p-spin models in 3 dimensions \cite{alvarez96}, for the 3d Ising spin
glass \cite{marinari} and a string in a disordered
medium~\cite{yoshino98}, confirming in each case the general features
of mean-field predictions.  In this Letter, we show that an accurate
determination of $x(C)$ for a structural glass model is indeed
possible, using standard simulation techniques.

The system we study is a 80:20 mixture of 1000 Lennard-Jones particles,
with interaction parameters that prevent crystallization~\cite{kob}.
In the following, we shall use as length, energy and time units the
standard Lennard-Jones units $\sigma_{AA}$ (particle diameter),
$\epsilon_{AA}$ (interaction energy), and $\tau = (m_A\sigma_{AA}^2/48
\epsilon_{AA})^{1/2}$, where $m_A$ is the particle mass and the
subscript $A$ refers to the majority species~\cite{kob}. The system has
been described in detail elsewhere, and its equilibrium (high
temperature) properties have been fully characterized. At the reduced
density $\rho=1.2$, a ``computer glass transition" is found in the
vicinity of $T=0.435$ and the slowing down of the dynamics seems to be
described well by Mode-Coupling theory~\cite{kob}. A first study of the
aging behavior of the correlation functions at low temperatures has
also been published recently \cite{kob_barrat}.

In order to obtain a fluctuation dissipation ratio, we need to compute
$C$ and $R$ for the same observable. Previous work~\cite{kob_barrat} focused
on the aging behavior of the incoherent scattering function for
wavevector ${\bf k}$:
\begin{equation}
C_k(t_w+\tau,t_w)= {1\over N} \sum_j e^{i{\bf k}\cdot 
\left({\bf r}_j(t_w+\tau)-{\bf r}_j(t_w)\right)} .
\end{equation}
In order to compute the associated response function, we use the
following numerical approach. A fictive ``charge" $\epsilon=\pm 1$ is
assigned randomly to each particle. An additional term of the form
$\sum_j \epsilon_j V({\bf r_j})$, where $V({\bf r})  = V_0\cos ({\bf
k}\cdot{\bf r})$ is a small ($V_0 <k_B T)$ external potential, is then
added to the  Hamiltonian. It is then easy to show that, {\it if one
averages over several realizations of the random charge distribution},
the time-correlation function of the observable $A_k = \sum_j
\epsilon_j \exp(i{\bf k}\cdot r_j(t))$ is the incoherent scattering
function. The procedure to generate the response function associated to
$C_k$ is thus straightforward: For a given realisation of the random
charge distribution, the system is  equilibrated at a high temperature
($T=5.0$), and quenched at $t=0$ to the desired final temperature
$T_f$. The evolution is followed with the field off until a waiting
time $t_w$, then the field is switched on and the response
$A_k(t_w+\tau,t_w)$ is monitored. The same procedure is repeated for
several (7 to 10) realisations of the charge distribution, in order to
get the response function. The quantity we obtain by this procedure is
then an integrated response function $M(t_w+\tau,t_w)$, defined as:
\begin{eqnarray}
\langle A_k(t_w+\tau, t_w)\rangle &=  &V_0 \int_{t_w}^{t_w+\tau} R(t_w+\tau, t) dt \\
& = & V_0 M(t_w+\tau,t_w) .
\end{eqnarray}

This procedure was carried out for three different values of the final
temperature $T_f$, namely $T_f=0.4, T_f=0.3$ and $T_f=0.1$.  The
amplitude of the external potential was chosen in such a way that a
linear response is obtained at each temperature. For $T_f=0.4$,
$V_0=0.2$ while for $T_f=0.1$ $V_0=0.05$. The wavevector was  $k=7.25$,
the location of the main peak in the structure factor. The runs had a
length of $5 \cdot 10^6$ time steps, corresponding to 100,000 time
units.

\begin{figure}
\psfig{file=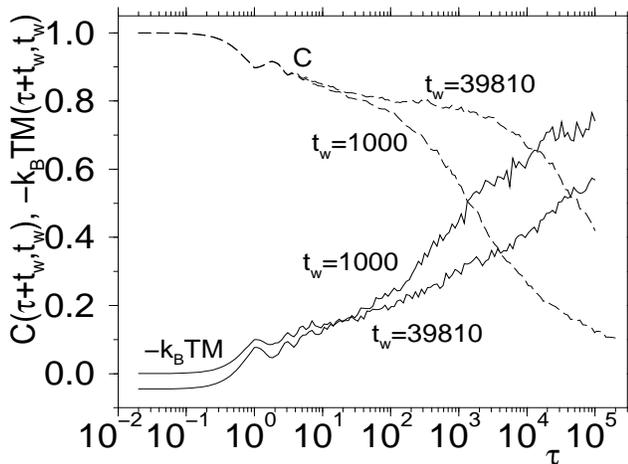,width=8.5cm,height=6.5cm}
\caption{Correlation function $C(t_w+\tau,t_w)$ (dashed lines) and
integrated response function $M(t_w+\tau,t_w)$ (solid lines) for
$T_f=0.4$, $k=7.25$, and two different waiting times.}
\label{MC_vs_t}
\end{figure}

\begin{figure}
\psfig{file=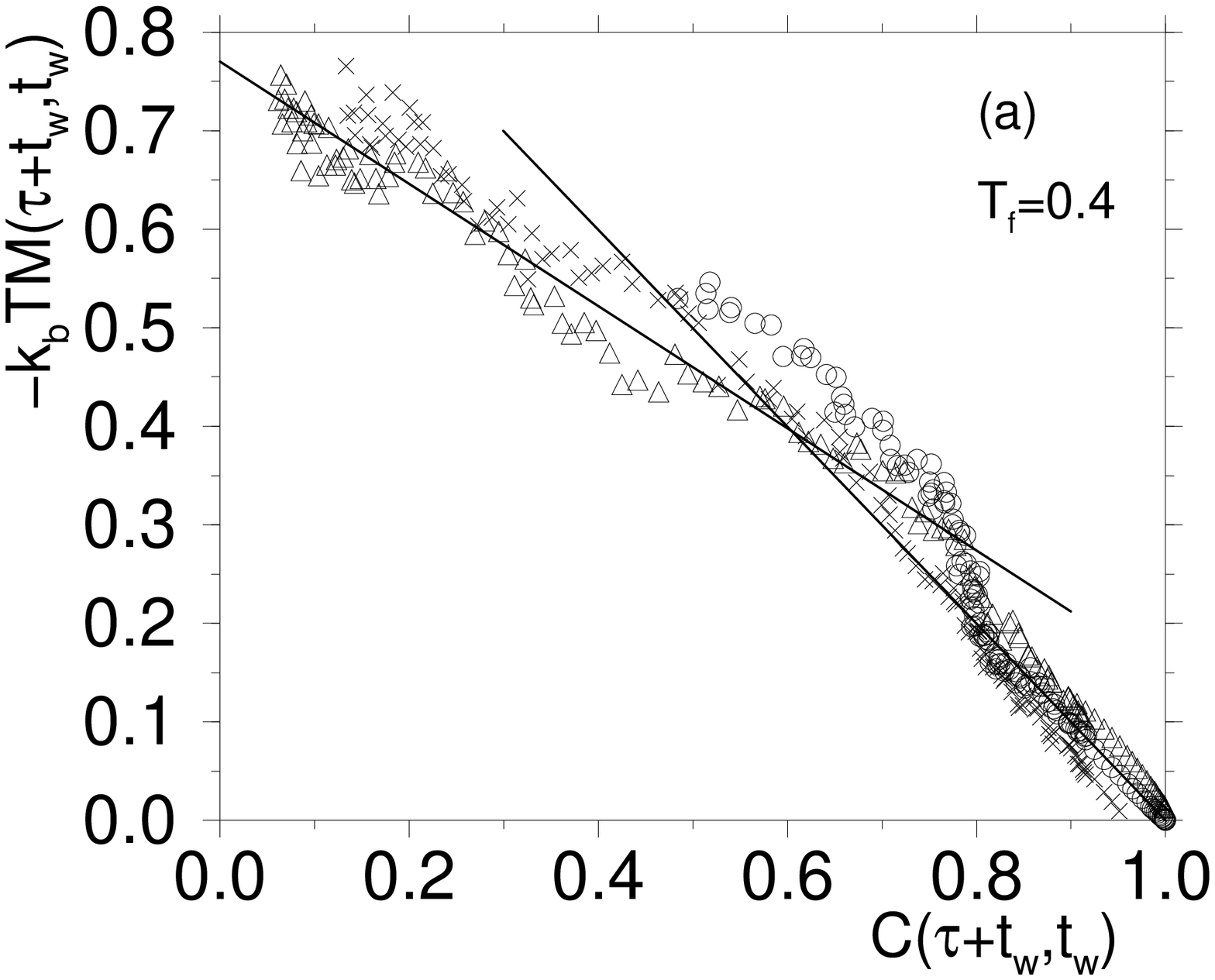,width=8.5cm,height=6.5cm}
\vspace*{2mm}
\par\noindent
\psfig{file=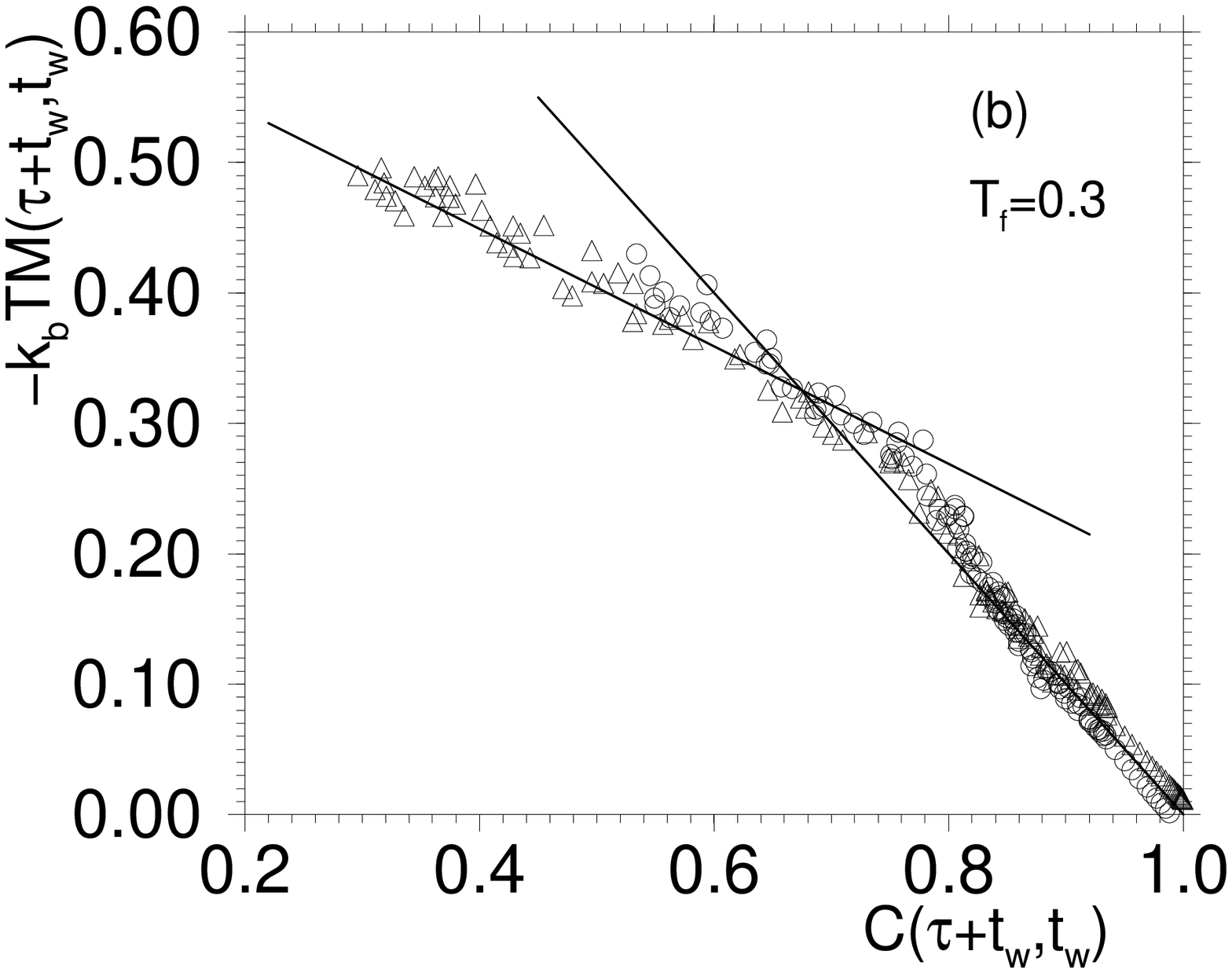,width=8.5cm,height=6.5cm}
\caption{Parametric plot of the integrated response function
$M(t_w+\tau,t_w)$ and the correlation function $C(t_w+\tau,t_w)$ for
$k=7.25$. a) $T_f=0.4$, Triangles: $t_w=100$. Crosses: $t_w=1000$.
Circles:  $t_w=39810$. The two straight lines have slopes $-1.0$ and
$-0.62$. b) $T_f=0.3$,  $t_w=1000$. Circles: $t_w=10000$. The straight
lines have slopes $-1.0$ and $-0..45$.}

\label{M_vs_C}
\end{figure}

Typical data for the integrated response and for the correlation
function is shown in Fig.~\ref{MC_vs_t}, for $T_f=0.4$ and $k=7.25$.
The way by which the correlation function and the integrated response
are related to each other can be understood well by means of a
parametric plot of $M$ versus $C$, as shown in Fig.~\ref{M_vs_C} for
different values of $t_w$ and two different $T_f$. If the generalized
fluctuation theorem holds, it is easily checked that $M$ can be written
as a function of $C$, with
\begin{equation}
M(C) =-  {1\over k_BT}\int_C^1 x(c) dc
\end{equation}
From Fig.~\ref{M_vs_C}, it is clearly seen that the usual FDT with
$x=1$ is very well verified for short times, i.e. values
of $C$ close to $1$, in that the curves are linear and have slope
$-1$.  For longer times a break in the curves is observed, i.e. the FDT
is violated.  Some transient effects are perceptible for the shorter
waiting time, but they tend to disappear with increasing $t_w$.  This
violation is compatible with the Ansatz (\ref{xC}),  since the
parametric curves obtained for different waiting times superimpose
satisfactorily.

For the regime in which the FDT is violated, the $M(C)$ curve can have
different forms~\cite{Bouchaud97,rieger}: Domain growth models predict
that $M(C)$ is a constant, whereas mean field models predict a linear
dependence, for the case of ``one step" replica symmetry breaking, and
a more general dependence for the case of continous replica symmetry
breaking.

As can be seen from Fig.~\ref{M_vs_C}, a good fit to the resulting $M$
versus $C$ curve is obtained with a piecewise linear function, which
corresponds to a piecewise constant $x(C)$:
\begin{equation}
x(C) = 1\  {\rm for } \ C > q_b, \quad x(C) = m < 1 \  {\rm for } 
\ C < q_b ,
\end{equation}
where $q_b$ is the value of $C$ at which the mentioned break in the
curves is observed.  Such a dependence has, e.g., been found for
mean-field ``p-spin'' models~\cite{Bouchaud97}. Thus our results give
support to the hypothesis first formulated by Kirkpatrick and coworkers
\cite{Kirkpatrick} and revived by Parisi \cite{parisi_jpa}, that
structural glasses belong to the same ``universality class" as
mean-field p-spin models.

From Fig.~\ref{M_vs_C}a we can read off $m\simeq 0.62$ and $q_b\simeq
0.6$. The latter value is clearly smaller than  the plateau value for
the correlation function in Fig.~\ref{MC_vs_t}, $q_{EA}\simeq 0.8$.
This means that the FDT appears to hold even for times at  which the
system is no longer time translationally invariant, a feature which is
not predicted by current theories of aging.

Finally, the dependence of $x$ on the final temperature can be
investigated. To explore this dependence we have also done simulations
at $T_f=0.3$ and $T_f=0.1$. In all cases, we  find that the $M$ vs. $C$
plot can  be fitted well by two straight lines.  Our results for $m$ as
a function of $T$ are thus given by: $T_f=0.4$, $m=0.62 \pm 0.05$;
$T_f=0.3$, $m=0.45\pm 0.05$; $T_f=0.1$, $m=0.2\pm 0.1$  Within the
accuracy of our data these values of $m$ are compatible with a linear
dependence on $T_f$, quite similar to that found by
Parisi~\cite{parisi_prl} for a soft-sphere system.
Such a linear dependance ($m(T_f)\sim T_f$) corresponds 
to a constant   "fluctuation dissipation effective temperature"  $T_{eff}=T_f/m$. 
The later concept,  introduced in \cite{CKP},  could help
rationalize the older "fictive temperature"
idea.

 We mention that, in his
analysis of the fluctuation dissipation relation, taking as an
observable the mean squared displacement, Parisi found that $m(T_f)$
can be approximated by $m(T_f)= T_f/T_c$ for $T_f<T_c$, where $T_c$ is
the ``mode coupling critical temperature" of the system under study. In
our case, $T_c\simeq 0.435$ \cite{kob}, so that our data is  in
contradiction with such a simple dependence of $m$ on $T$. Our results
are much more similar to the ones found by Alvarez {\it et al.} for the
p-spin model in that these authors found for a temperature a bit below
$T_c$ a value of $m$ which is significantly smaller than 1.  The reason
for this difference might be related to the much smaller waiting times
used in Ref. \cite{parisi_prl}\cite{note}.
In any case, it is not clear why the mode coupling critical temperature
should play a particular role in the present analysis. If the same type
of simulations would be carried out at a temperature slightly above
$T_c$, we expect that interrupted aging will be observed, so that at
short $t_w$ a violation of FDT occurs. As $t_w$ increases, equilibrium
will progressively be approached, and the $M$ versus $C$ plot will
approach a straight line with slope $-1$. Hence the main difference
between $T>T_c$ and $T<T_c$ will be that for $T>T_c$ the $M$ versus $C$
plot depends on $t_w$, as has been shown for the
Sherrington-Kirkpatrick model in three dimensions~\cite{franzrieger}.
However, for $T$ close to, but above, $T_c$ this $t_w$ dependence will
be so weak that it can be neglected for all practical purposes. In terms
of the "effective temperature" $T_{eff}=T_f/m$, our system falls 
out of equilibrium above $T_c$, so that we can expect $T_{eff}$ to be larger
than $T_c$ - which is indeed the case.

In summary, we have shown that the fluctuation dissipation ratio of a
supercooled liquid out of equilibrium can be computed with good
accuracy from MD simulations. Several nontrivial features predicted by
the theory of mean-field spin glasses, beginning with the existence of
a waiting time independent  function $x(C)$,  seem to be present also
in the model structural glass we study.  Our data is compatible with a
stepwise constant $x(C)$, which would correspond to a phase space
structure similar to that of spin systems undergoing one step replica
symmetry breaking. This means that phase space is divided by high
barriers into different valleys each of which has the same statistical
properties. (The case of continuous replica symmetry breaking
corresponds to a case that the valleys are organized in a hierachical
way.). In any case, finding a nonzero value of $m$ seems to be a clear
indication that a ``domain growth" picture is not applicable to our
model.  A quantitative comparison between theoretical predictions and
simulation results, similar to the work carried out for testing mode
coupling theory, would be required in order to fully clarify the
situation with respect to nonequilibrium dynamics.  However,
unfortunately such calculations are currently not yet feasible.

\acknowledgments 
We benefited from useful discussions with L. Cugliandolo and J. Kurchan
and correspondence with H. Rieger.  We thank G. Parisi for pointing out
several inconsistencies in a previous version of this work.
This work was supported by the Pole
Scientifique de Mod\'elisation Num\'erique at ENS-Lyon, and the
Deutsche Forschungsgemeinschaft through SFB 262.

\vspace*{-7mm}
\par\noindent


\begin{references}
\bibitem{note} The longest waiting time considered by Parisi are of the
order of $10^4$ Monte-Carlo steps, which would typically correspond to
$100$ in Lennard-Jones units.  Other differences involve smaller system
sizes, and quenching directly from $T=\infty$ to $T_f$.
\end{references}
\end{document}